\documentclass[12pt,preprint]{aastex}

\slugcomment{Submitted to ApJ April 24, 2009}

\shorttitle{Wide Companions to HR8799}
\shortauthors{Close \& Males}

\begin{document}


\title{A Search for Wide Companions to the Extrasolar Planetary System HR 8799}


\author{Laird M. Close and Jared R. Males}
\affil{Steward Observatory, University of Arizona, Tucson, AZ 85721}








\begin{abstract}
The extrasolar planetary system around HR 8799 is the first
multiplanet system ever imaged. It is also, by a wide margin, the
highest mass system with $>27$ Jupiters of planetary mass past 25
AU. This is a remarkable system with no analogue with any other known
planetary system. In the first part of this paper we investigate the
nature of two faint objects imaged near the system. These objects are
considerably fainter (H=20.4, and 21.6 mag) and more distant
(projected separations of 612, and 534 AU) than the three known
planetary companions b, c, and d (68-24 AU). It is possible that these
two objects could be lower mass planets (of mass $\sim5$ and $\sim3$
$M_{Jup}$) that have been scattered to wider orbits. We make the first
direct comparison of newly reduced archival Gemini adaptive optics images to
archival HST/NICMOS images. With nearly a decade between these epochs
we can accurately assess the proper motion nature of each candidate
companion. We find that both objects are unbound to HR 8799 and are
background. We estimate that HR 8799 has no companions of H$<22$ from
$\sim5-15$\arcsec. Any scattered giant planets in the HR 8799 system
are $>600$ AU or less than $3$ $M_{Jup}$ in mass. In the second part
of this paper we carry out a search for wider common proper motion
objects. While we identify no bound companions to HR 8799, our search
yields 16 objects within 1 degree in the NOMAD catalog and POSS DSS
images with similar ($\pm20$ mas/yr) proper motions to HR 8799, three
of which warrant follow-up observations.
\end{abstract}





\section{Introduction}

There have been several surveys to directly image extrasolar planets
for the ground with adaptive optics (AO) and from space with
HST. Until very recently all surveys returned null results, and so it
was generally assumed that wide, massive, extrasolar planets would be
rare (at least around Sun-like stars) past 20 AU (\citealp{laf07}; \citealp{nie08} and references within).
However, in November 2008 \citealp{mar08} announced the discovery of 3
planets orbiting the A5V star \object{HR 8799}, based on Near-IR
imaging at the Keck and Gemini telescopes.  Using data from Keck in
2004-2008 and Gemini in 2007-2008, they were able to establish
common-proper motions (assuming the non-common motion component of
$\sim25$ mas/yr for b and c was due to orbital motion of the planets
around the star).  The three planets, HR 8799b,c, and d orbit at
approximately 68, 38 and 24 AU respectively. \citealp{mar08} estimate
effective temperatures of 870, 1090, and 1090K for the three planets,
and arrive at estimates for mass of 7, 10, and 10 $M_{Jup}$.  These
estimates are based on the estimated 30-160 Myr age of HR 8799 and
hybrid theoretical cooling tracks (luminosity vs. age) for giant
planets \citep{mar08}.

It is possible that additional companions could be discovered during a
search at wider separations.  The binary fraction of late A stars is
at least 70\% \cite{bat09}.  Recently, \cite{ver09} have predicted a
population of giant planets at large separations (100 AU - 10,000 AU)
from stars hosting relatively close-in planets, where the distant
objects are dynamically scattered/pumped to large separations after
system assembly. Such effects may explain the very wide, low mass,
companions GQ Lup b at $>100$ AU \citep{neu05} and/or AB Pic B at
$>250$ AU \citep{cha05}.  There is certainly the possible existence of
low mass, reddened, stellar companions that have not yet been
detected. It is even possible that HR 8799 has a small, common-proper
motion group, around it. Given the unique nature of the HR 8799
system, a search for wide ($>100$ AU) companions is important and
motivated this paper.

\section{Observations \& Reductions}

\begin{figure}
\epsscale{.80}
\plotone{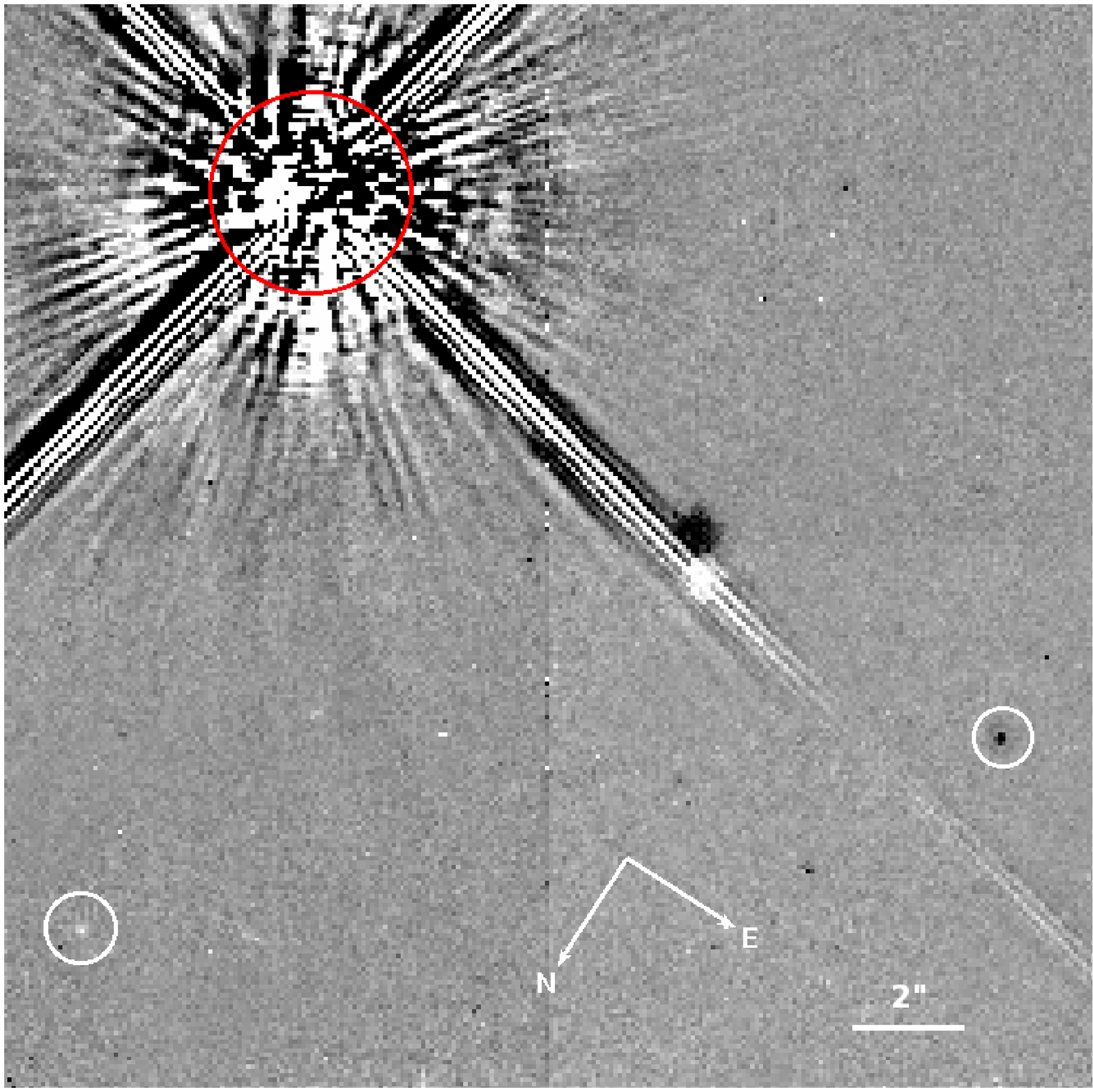}
\caption{Roll subtracted NICMOS F160W image from 1998, showing the 2
    candidate companions ``C'' (to the left) and ``B'' (to the right)
    first identified by \cite{low05} -- each is circled in white.
    Each was visible in only one roll. The red circle around the star
    shows the 1.7\arcsec radius of HR 8799b. The NE compass is correct for the positive image (HD 8799B), however for HR 8799C (negative image) the compass should be rotated by $30^{o}$ counterclockwise \label{fig1}}
\end{figure}

\begin{figure}
\epsscale{1.0}
\plotone{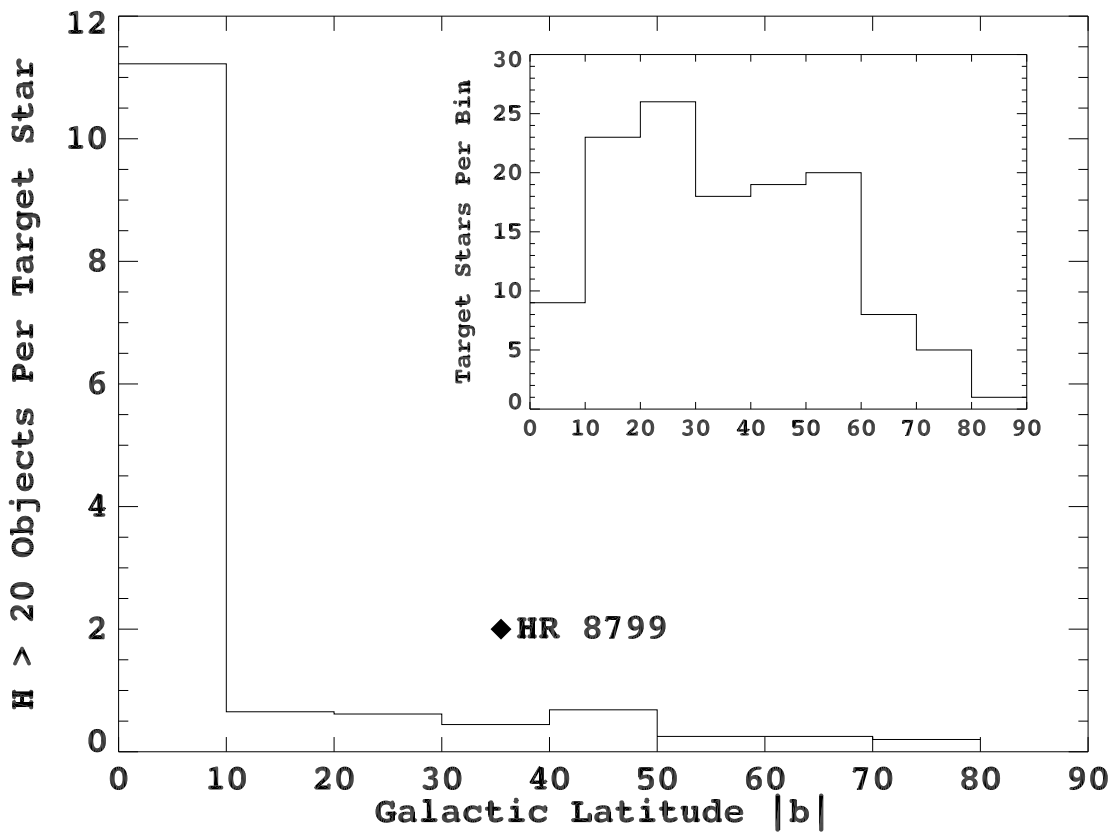}
\caption{Histogram of the $H > 20$ point source detections made by
the HST/NICMOS survey \cite{low05} and Gemini Deep Planet Survey \cite{laf07},
combined, binned by galactic latitude.  These results demonstrate the
low density of such background objects away from the galactic
plane. We note that HR 8799's ``overdensity'' of such companions hints
that there might be a co-moving group (or at least a real companion)
around HR 8799.  The inset shows the total number of stars observed in
each latitude bin by the two surveys.\label{fig3}}
\end{figure}

On Oct. 30 1998 UT HST/NICMOS observed HR 8799 with its coronagraph,
and 2 candidate companions were identified \cite{low05} in the roll
subtracted images.  These faint point sources where reported at
13.7\arcsec and 15.7\arcsec (540 AU and 619 AU) with H magnitudes of 21.6 and 20.4,
respectively. These are much wide separations
 than the 3 confirmed planets which were not discovered in the NICMOS
data at the time \citep{laf09}.  Figure \ref{fig1} shows our own roll subtraction of the
otherwise already pipeline reduced NICMOS data with the 2 objects
identified. HR 8799's high galactic latitude ($b =-35^{o}$) indicates
that there is a finite chance one (or both) are not background objects
(see Fig \ref{fig3}). However, we need to determine if these are
real common proper motion companions or simply background objects. For lack of a better nomenclature
we have elected to call the closest candidate ``HR 8799B'' and the
farther one ``HR 8799C'', however, the use of these labels does not
imply they are physical companions.

In October 25, 2007 \cite{mar08} obtained 118x30s Angular Differential
Imaging (ADI; \cite{mar06}) dataset of HR 8799 on the Gemini-North
telescope with the Altair AO system and NIRI NIR camera. Typically an
ADI dataset would not be the ideal method of imaging faint companions
at 13.7\arcsec and 15.7\arcsec since there will be a risk of
rotational blurring of the images in the azimuthal direction. However,
in this HR 8799 dataset the integrations of 30s were short enough that
only minimal azimuthal blurring occurred --and only then in the
fastest rotating images near transit. Therefore, we were able to
create a new custom ``ADI-like'' IRAF pipeline to produce ``Wide
Field'' ADI images (WiFi ADI) to image faint companions at the very
edge of ADI datasets. Since the focus of WiFi ADI is wide companions
there is no need to use more advanced ADI reductions like LOCI
\citep{laf07b}.

Our WiFi ADI pipeline is very similar to standard ADI reduction and
runs in a standard IRAF environment. In ADI the telescope rotator is disabled
and so the median of the images gives an estimate of the master PSF
without ``contamination'' from real objects on the sky. Then one must
subtract the master PSF off each individual frame after a cross
correlation alignment of each frame as in \citealp{clo02}. Once the
frames have the PSF removed they need to be rotated by the parallactic
angle and median combined (the subroutine to calculate the rotation
angle is a custom script developed for Gemini data in
\citealp{clo03}). One way that the WiFi ADI pipeline differs from the
standard is that it is optimized to preserve any faint off-axis
objects that might fall past the {\it edge} of the IR array in the
majority of the individual ADI frames. For example, the code
accurately masks many bad pixels in the corners of the NIRI array
and uses a final median combine of all 118 re-rotated (master PSF
subtracted) images with no pixel clipping or rejection (so even the corner pixels
of the individual frames are utilized in the final WiFi image). We also
carefully offset each image by the mode of the outer region of each
image, this allows the final median combine to be most sensitive in
the outer regions of the image.

\begin{figure}
\epsscale{.80}
\plotone{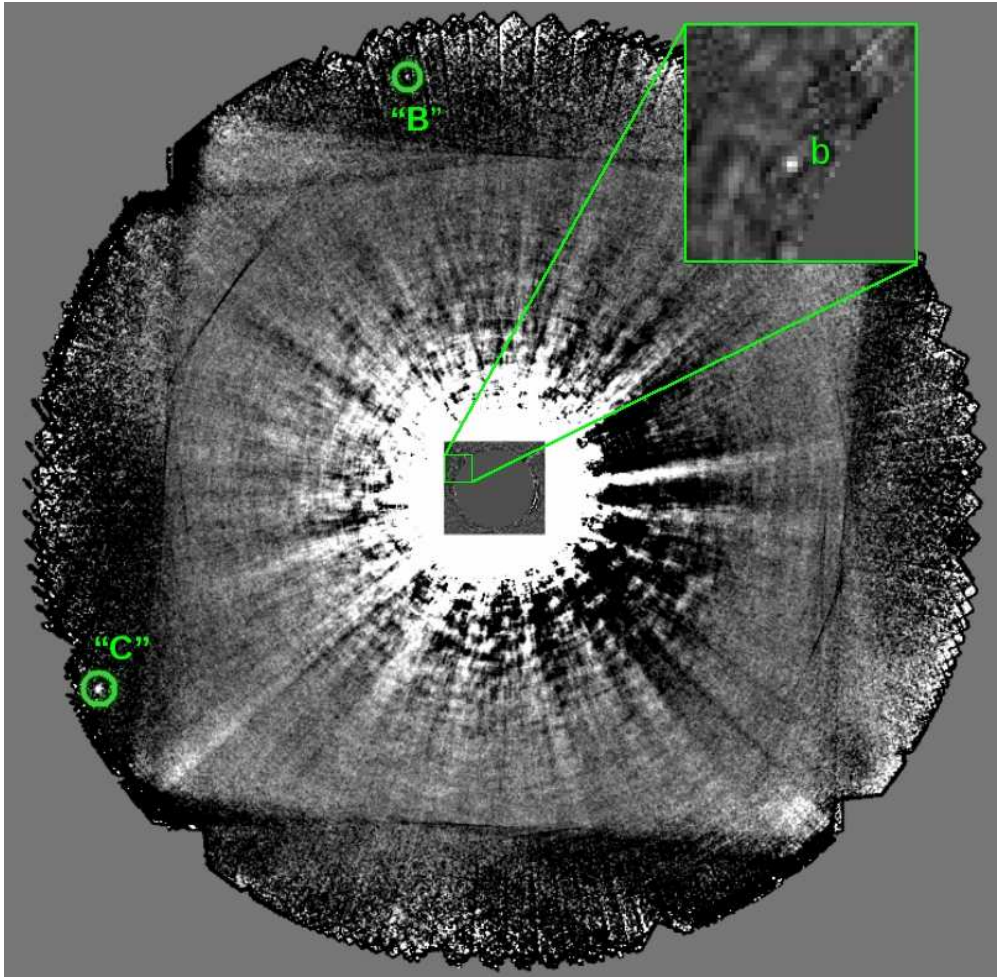}
\caption{Our WiFi ADI reduction of the Oct 25, 2007 Gemini Altair/NIRI
dataset of \cite{mar08}. Note the locations of two very faint ``B''
($PA=13^{o}$; Sep=13.78\arcsec) and ``C'' ($PA=115^{o}$;
Sep=14.86\arcsec) sources. North is up East to the left. Also there is
a detection of the planet HR 8799b which is located near its nominal
position reported in Keck AO images \citep{mar08}. For clarity we have
inserted a zoomed in box centered on HR 8799b, the other planets (c \& d)
were too close and faint to be clearly detected in this WiFi ADI reduction.
\label{fig2}}
\end{figure}

Our WiFi ADI pipeline when used with the NIRI detector with its
0.0219''/pixel scale is then capable of creating a round WiFi ADI FOV
of 31.71\arcsec in diameter when there is at least 90
degrees of field rotation during the ADI observation, and the
object is centered in the detector. In the case of the Oct. 25 2007 Gemini HR
8799 data of \citealp{mar08} the above assumptions are all true and our
pipeline produced a final image (see Fig. \ref{fig2}) of all $H<22$ objects within
$\sim5-15\arcsec$ of HR 8799.

\section{Analysis}

As is clear from figure \ref{fig1} and \ref{fig2} there are two faint
 objects near HR 8799.  Based on detailed search of the literature and
 the VLT/Gemini/Subaru/HST archives we have concluded that there has
 not been any published attempt to recover ``B'' or ``C'' until
 now. HR 8799A has a total proper motion of 119 mas/year so the
 recovery of the two candidate companions at the same separations as
 in 1988 wrt HR 8799A would be an unambiguous confirmation of their
 physical association.  These objects are both fainter than the
 HD8799b-d planets (they would have masses of 3 and 5 $M_{jup}$ on the
 0.1 Gyr \cite{bar02} COND tracks; which predict reasonable
 luminosities at these ages for higher mass objects;
 \citealp{clo07b}), consistent with being scattered by the heavier,
 close-in, planets.

\begin{figure}
\epsscale{1.00}
\plotone{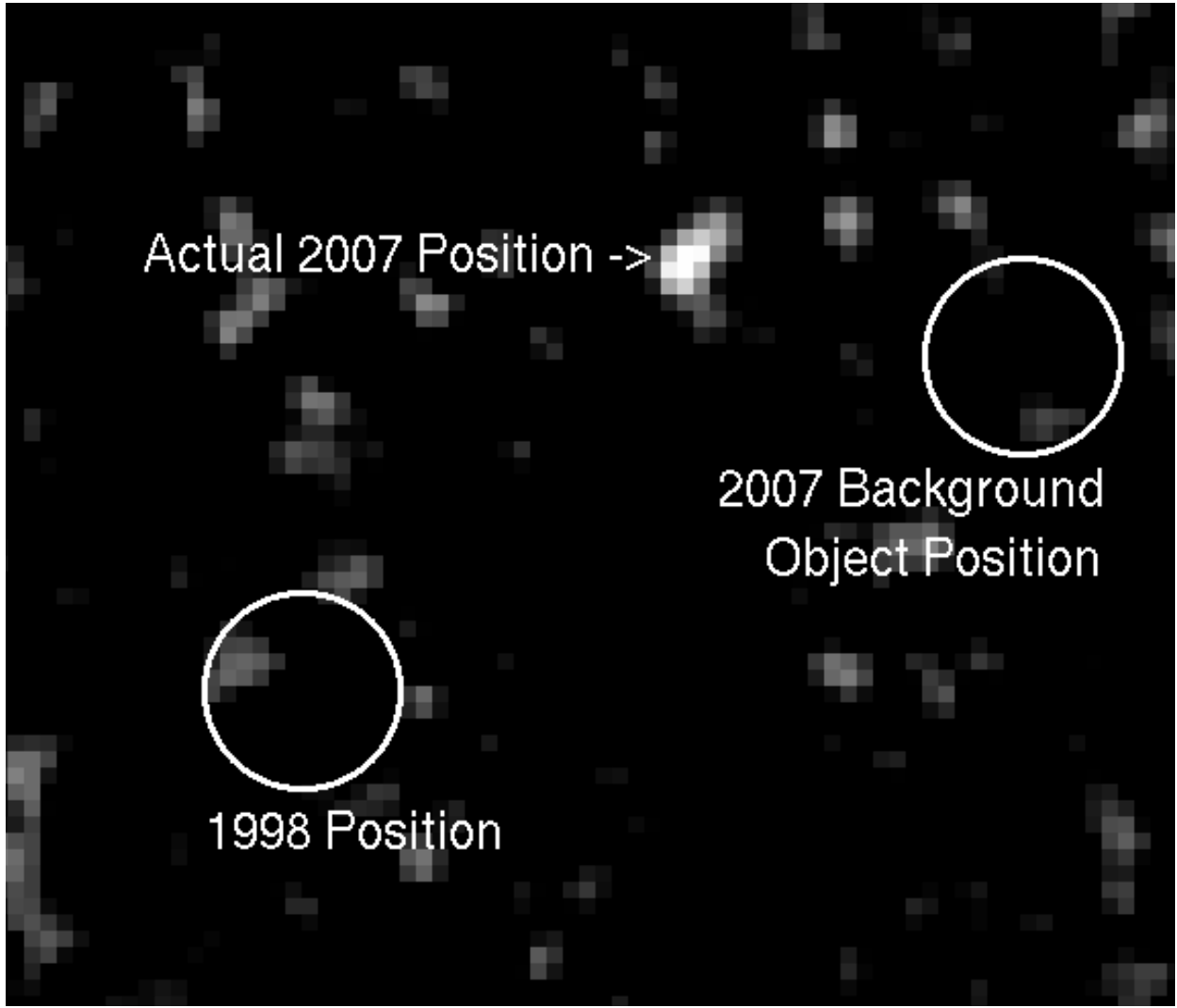}
\caption{A 1.5x1.2\arcsec section of our reduced WiFi image of the Gemini 2007 Oct 25
dataset. Note that the H=21.6 object ``B'' detected by
HST/NICMOS is clearly not common proper motion with HR 8799A. The size
of the error circles is dominated by 0.5\% platescale uncertainties
across the field. North up East left,
$0.0219\pm0.0001\arcsec$/pix. The object is slightly elongated due to
the significant field rotation 14.15\arcsec off axis during these exposures near transit.
\label{figADIa}}
\end{figure}

\begin{figure}
\epsscale{1.0}
\plotone{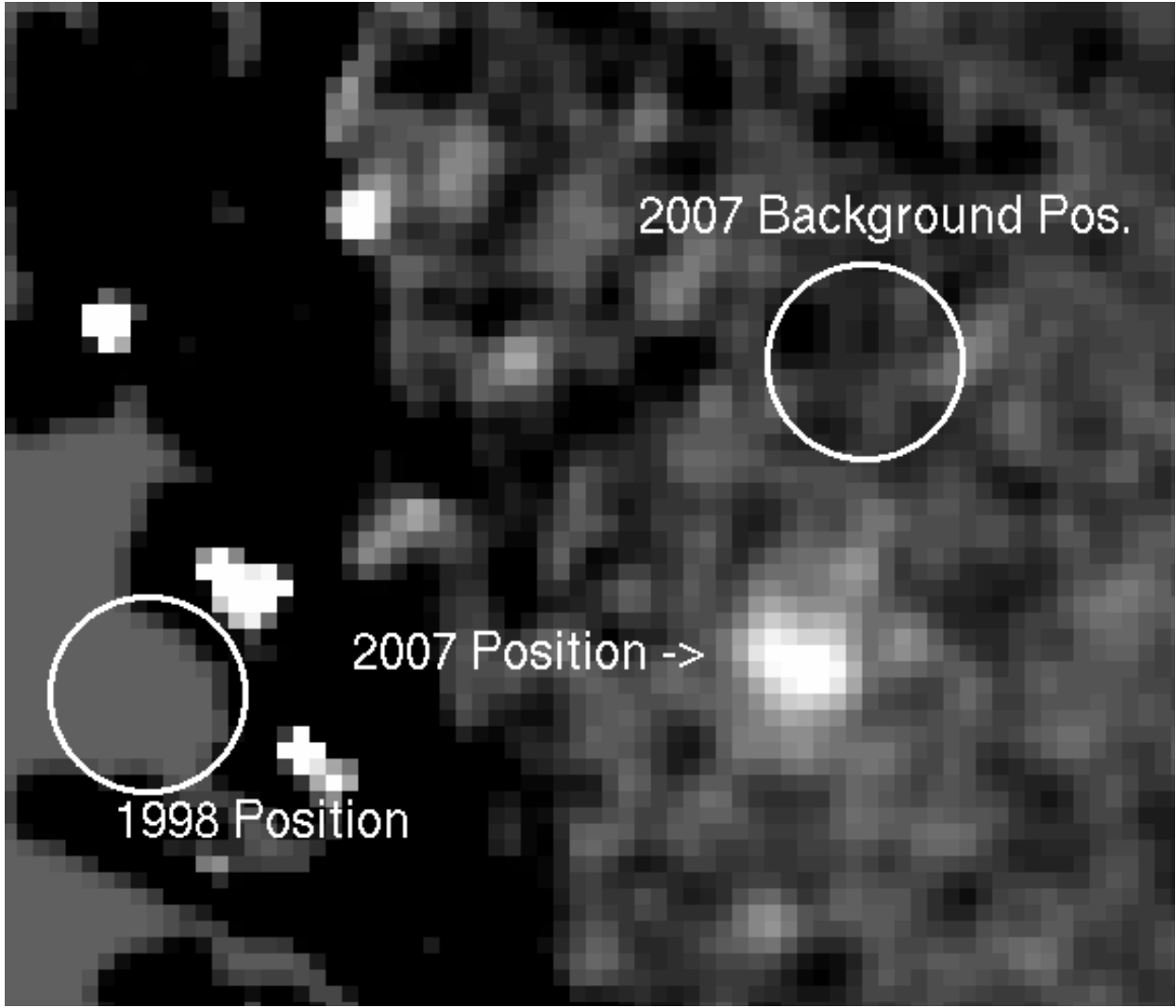}
\caption{Same as fig \ref{figADIa} but for the H=20.4 object ``C''. As
with ``B'' the ``C'' object also appears to be background as
well. Note that the bright small pixels ``clumps'' are bad pixels at
the very edge of the array, there is only one real object in the image
(noted by the arrow).
\label{figADIb}}
\end{figure}

In Figure \ref{figADIa} and \ref{figADIb} the 2007 positions of these
two faint companions are shown. In both cases the current positions
are much closer to the locations calculated for distant background
objects rather than physical companions. Therefore, our astrometry
proves that the NICMOS companions of \citealp{low05} appear to be faint
background objects unrelated to HR 8799A (see Table \ref{tbl-1} for a detailed
list of our astrometric measurements).


\subsection{Does HR 8799b Show any Parallax Motion of a Background Object?}

\begin{figure}
\epsscale{1.2}
\plottwo{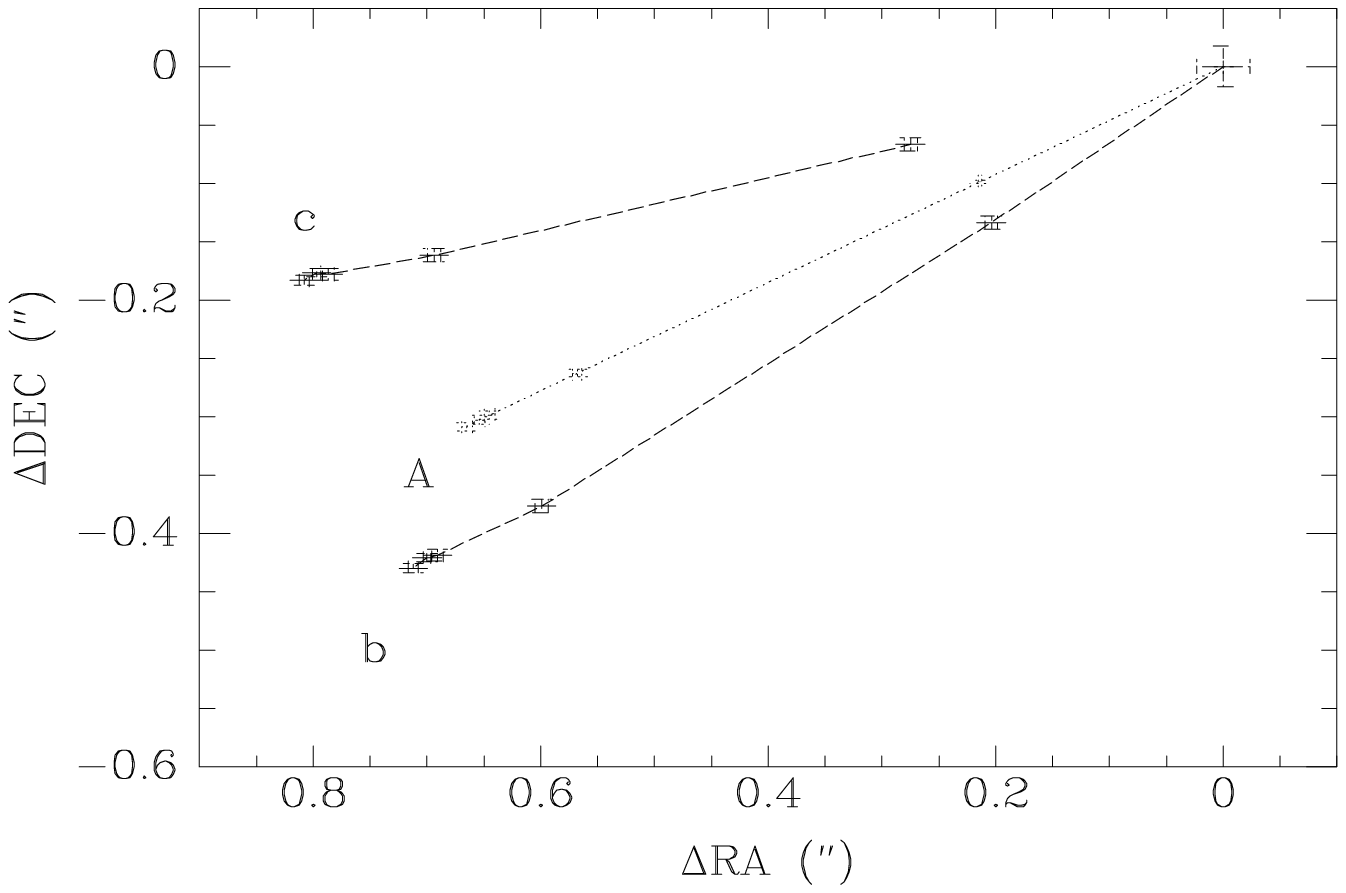}{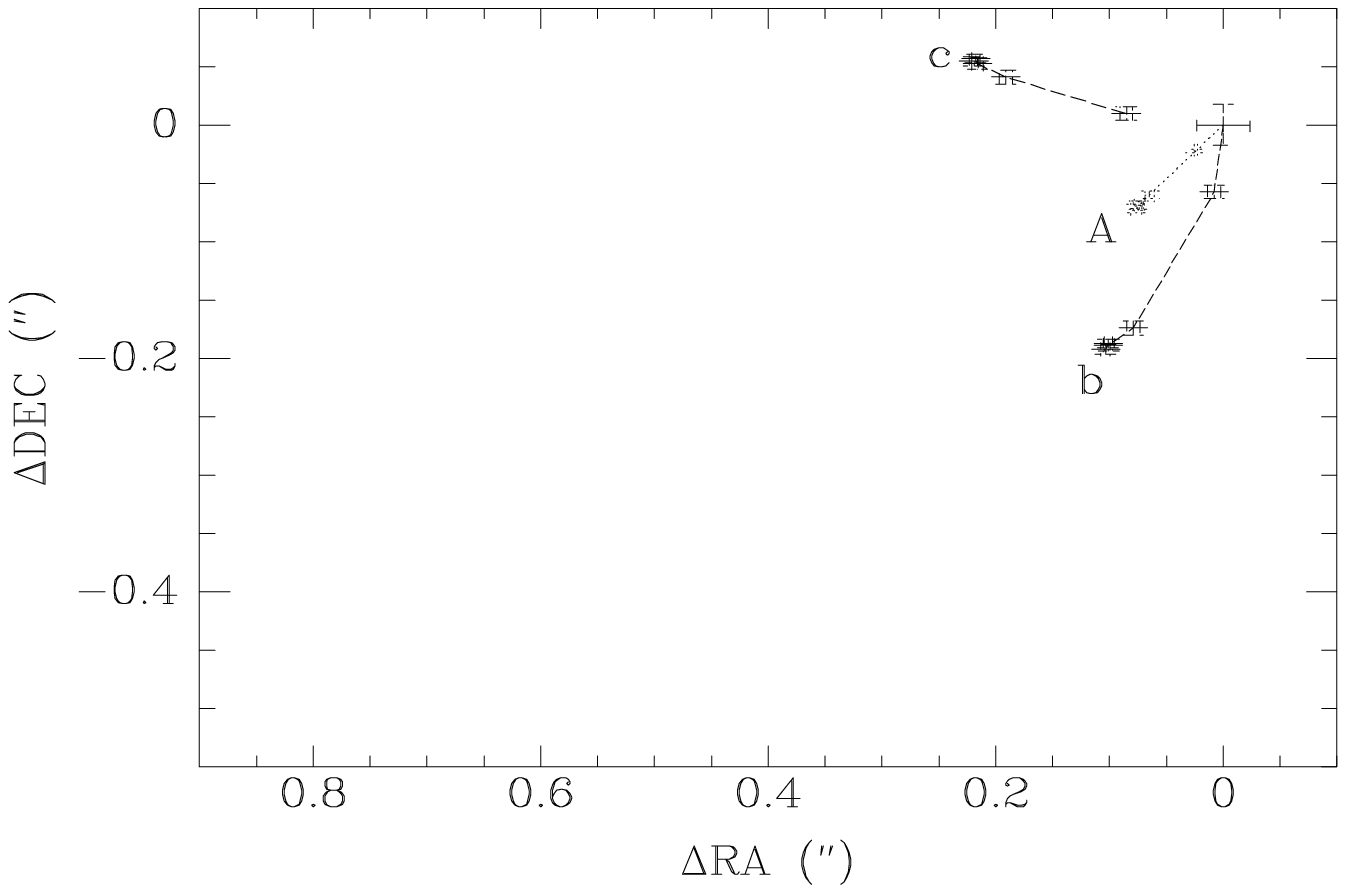}
\caption{{\bf Left:} The proper motions of HR 8799A, and extrasolar planets candidates HR
8799b and HR 8799c  plotted over a 6.14 yr period --normalized to a common starting point in 2002 of the \cite{fuk09} observation.
{\bf Right:} Same as to the Left but the Solar reflex motion of
  95.51mas/yr in RA and -38.51mas/yr DEC at the position of HR 8799A
  (d=39.9pc) has been removed. Hence, this is a plot of the true motions of HR 8799A, and extrasolar planets candidates HR 8799b
  and HR 8799c (all wrt the LSR). It is interesting to note that HR 8799A, b, and c appear to have clearly 
different proper motions once the Solar motion is subtracted. These differences in motion are most likely due to
orbital motion of b and c around A \citep{mar08}.
\label{fig4}}
\end{figure}

In the direction of Pegasus most nearby stars appear to be moving
towards the East South East. This is due to the Sun's motion in the
opposite direction wrt the LSR. In fact, the Sun's space motion causes
a stationary object (wrt LSR) at 39.9 pc to have a measured proper
motion of 95.51 mas/yr to the East and 38.51 mas/yr to the South
--based on values for the Solar motion given by \citealp{jv93}. While
it is clear that planets HR 8799b and HR 8799c have similar proper
motions on the sky to HR 8799A (see left side of Fig. \ref{fig4}), once we
subtract the Solar motion there is less agreement (see right side of
Fig. \ref{fig4}). However, this lack of common proper motion can be explained
by increased velocity due to orbital motion of b and c around A
\citep{mar08}.

For the planet HR 8799b (which has the largest timeline of
observations), there appears to be some ``scatter'' in its measured
position from A from the nearly straight line expected for b's long
period orbit. The exact solution for a stable orbit of the massive planets b, c, and d is still somewhat uncertain 
\citep{fab08}. 

\begin{figure}
\epsscale{1.0}
\plotone{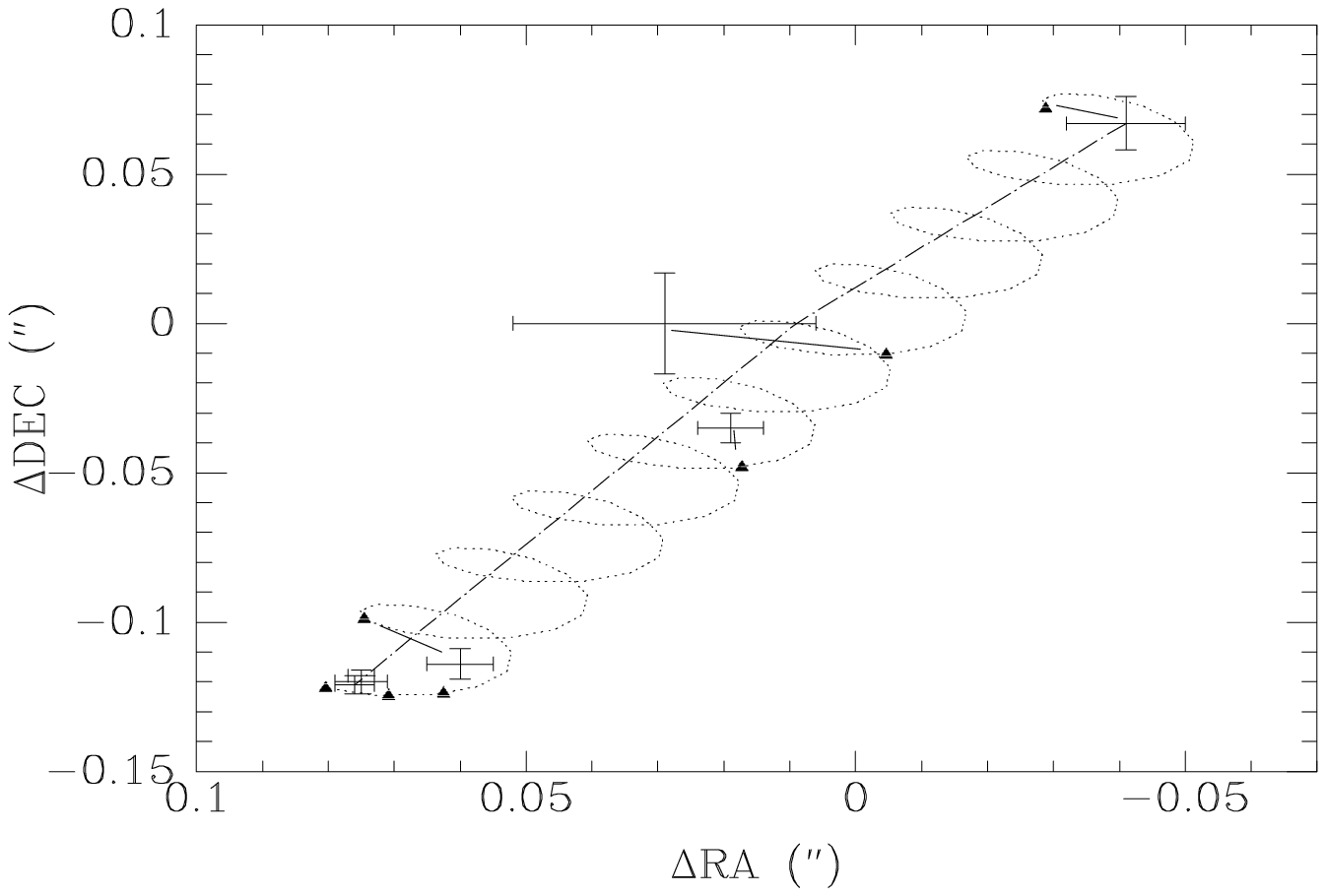}
\caption{Here we plot 1$\sigma$ errors of ten years of HR 8799b
relative astrometry of b wrt A from \citealp{fuk09} (Subaru AO; largest errors),
\citealp{laf09} (NICMOS; Oct 30 1998 top data point), and the rest of the data points from
\citealp{mar08} (Keck AO). The dotted line traces the ``reverse'' parallax
that should be observed if HR 8799b is a background object at 100
pc with a similar proper motion to A. The triangles denote time stamps to each observation date over the
9.885 yr period since the NICMOS observation. Note how neither a
slightly curving ``linear'' orbit or the 100 pc background object
model fits all data points simultaneously inside the 1$\sigma$
errorbars. However, the fit for the 100 pc model is quite poor, hence, we can
reject the hypothesis that this ``scatter'' in HR 8799b's position wrt
HR 8799A is due to the reverse parallax of a background object at 100
pc. We do not plot our new Gemini Oct 17, 2007 position for b since
there are unknown systematic errors in the data at the $\pm44$ mas
level, hence little statistical weight would be added by such a data point.
\label{fig5}}
\end{figure}

In Fig. \ref{fig5} we consider the question what would this
``scatter'' resemble if b was actually a background object at 100 pc
that had similar proper motions, by chance, to HR 8799A (at a distance
of 100 pc HR 8799b would roughly have the normal luminosity and colors of a
background L dwarf). In this model b's position should show a
``reverse'' parallax w.r.t. HR 8799A with an amplitude of 60\% that of
HR 8799A's parallax. We calculate parallax of HR8799 in the usual
manner \cite{bil07} and then multiply by -0.6 to calculate ``reverse''
parallax.

In Fig. \ref{fig5} we show two models for the nature of HR 8799b
motion: a background object at 100 pc; and a simple ``linear''
planetary arc. The small triangles in Fig. \ref{fig5} denote time
stamps in the 100 pc model to each observation date over the 9.885 yr
period since the detection of b in a LOCI analysis of the NICMOS
dataset by \citealp{laf09}. We note how neither the slightly curving
``linear'' orbit or the 100 pc background model fit all data points
simultaneously inside the 1$\sigma$ errorbars.  The ``linear'' orbit
is a simpler model with a better fit, but it certainly is not perfect
since the reduced of $\chi_{\nu}^2\sim2$ gives an $\sim8\%$ chance
that it is the correct model --assuming no unknown systematic errors
in the astrometry. On the other hand, a background source at 100 pc
model can be rejected with 99.95\% confidence
($\chi_{\nu}^2\sim4.4$). While there is still significant scatter in
the linear orbit fit for b, we cannot reject it (we also do not, at
this time, know what orbit to fit it to \citep{fab08}). However, we
can reject the hypothesis that this scatter in HR 8799b's position is
reverse parallax due to it being a background object at 100pc
(assuming, of course, the reported astrometric values and errors are
correct).

\subsection{A Search For Other Common Proper Motion Companions to HR 8799}

\begin{figure}
\epsscale{1.0}
\plotone{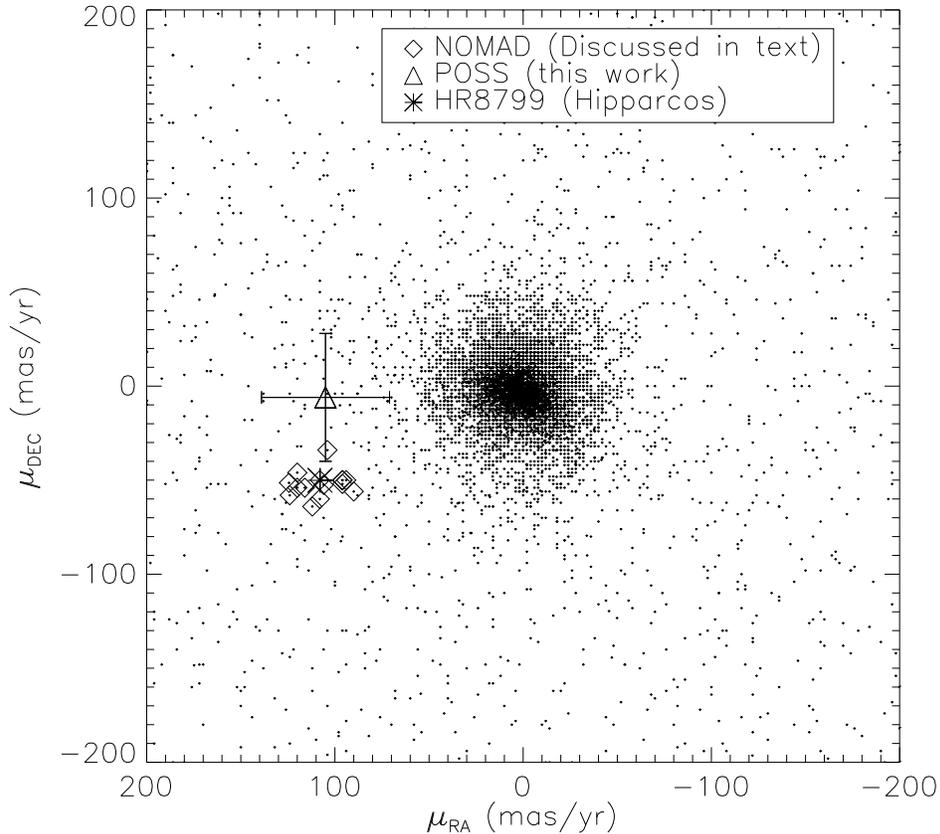}
\caption{ Proper motions of 12875 objects within 2 square degrees
centered on HR 8799A.  The NOMAD database was searched to recover all
objects in this field with proper motion measurements.  Diamonds
highlight the 15 objects with proper motion vectors within
$\pm20$mas/yr of HR 8799's vector (see Table \ref{tbl-2}).  We also
blinked the POSS1 (1951) and POSS2 (1991) red images to search for
stars with proper motions, yielding 1 interesting object (triangle)
with no proper motion measurement in NOMAD (see Table \ref{tbl-3}).
The error bars show our $\pm34$ mas/yr uncertainty for that one POSS
object some 280\arcsec from HR 8799A.
\label{fig7}}
\end{figure}

Even though the ``B'' and ``C'' objects are clearly background, there
exists the possibility of other wider companions to HR 8799A that
might be found by searching nearby for similar proper motion.  We
searched 12875 objects within $\pm1^{o}$ of HR 8799 in the the US
Naval Observatory Merged Astrometric Dataset (NOMAD) catalog
\citep{nomad} with non-zero proper motions.  See Figure \ref{fig7} for
a plot of the resulting proper motion vectors. Most objects inside
$1^{o}$ of HR 8799 are background and so do not have as large a proper
motion as HR 8799, which is dominated by the Sun's motion w.r.t the
LSR.  In Figure \ref{fig7} we have highlighted those objects which
have proper motion vectors within a circle of radius 20 mas/yr around
HR8799's proper motion.  This yields a list of 15 objects, two of which
may warrant follow up observations.  We further discuss these objects
below.

To increase our sensitivity to fainter objects that might have been
missed by previous surveys we manually examined the Palomar
Optical Sky Survey (POSS) POSS1 (1951 August 12 08:12:00 UT) and POSS2
(1991 October 02 05:47:00 UT) Red DSS images to search for objects within
1 square degree of HR 8799A with similar proper motions.
The POSS1 image has a plate scale of $1.7\arcsec$ per pixel, whereas
POSS2 has $1.0\arcsec$ per pixel.  We first magnified the POSS1 image
to match the POSS2 plate scale.  Next 25 background stars were
selected, and their positions measured by centroiding in each image.
We then compared their positions on the two images to determine the
optimum rotation ($\approx 0.7$ degrees) to apply to the POSS1 image.
We then selected an additional 25 background stars for a total of 50
stars spread across the images to build a background reference frame.
Proper motions were then measured for individual stars by determining
the average relative change in offset from the 50 background stars
between the 2 images

A different sample of 30 background stars was selected to test our
technique.  These were each compared to the 50 reference stars in
each frame, and their proper motions were calculated from the average
change in offset from the reference stars.  The overall average of
these measurements was $\mu_{RA}=0.81 \pm 3.50$ mas/yr and $\mu_{DEC}
= 4.92 \pm 3.04$ mas/yr.  We subtract these offsets from the proper
motion measurements made for individual stars.

For bright stars there is a $\pm0.3$ pixel uncertainty in centroid position
(estimated from $FWHM/\sqrt{SNR})$, and for the faintest objects that
we considered a $\pm0.95$ pixel uncertainty.  Given the 40.139 year
baseline between the two positions this amounts to $\pm10.6$ mas/yr of
proper motion uncertainty for bright stars, and $\pm33.5$ mas/yr for faint
objects.  We combine these in quadrature with the standard error of
our reference stars estimated from the 30 test stars above to obtain
an estimated uncertainty of $\approx \pm11$ mas/yr and $\approx \pm34$
mas/yr.  The bright star case compares favorably with other POSS based
proper motion surveys, e.g. \citealp{lep05} with $\approx \pm8$ mas/yr.

The $1^o$ X $1^o$ area of sky centered on HR8799 was searched manually
by blinking between the registered POSS1 and POSS2 images.  In order
to compensate for subtle changes in magnification and field
distortion, the image was re-centered on convenient background stars in
each small area being searched.  When a proper motion candidate was
identified, its position was measured by centroiding and then its
average change in offset relative to the 50 reference background stars
was calculated.

As a check on our measurements, we compared our measurements of proper
motion for 2 bright high proper motion stars in the field, NLTT 55870
and NLTT 55853, with POSS based proper motion measurements
\cite{lep05}.  We measured ($\mu_{RA}, \mu_{DEC}$) = (184.7, -43.5)
and (178.2, -61.4) respectively compared to the values of (185, -44)
and (164, -63) reported by \citealp{lep05}.  These measurements are
consistent with each other within our bright star uncertainty of $\approx \pm11$
mas/yr.

We note that we are comparing measurements of proper motion relative
  to a sample of background stars, which is not the true absolute
  proper motion.  \citealp{lep05} discuss this systematic effect and
  attribute it to bulk motion of the reference stars.  Taking NLTT
  55870 and NLTT 55853 as an example, they apply correction offsets of
  $\Delta\mu_{RA}=+4$ mas/yr and $\Delta\mu_{DEC} = -7$ mas/yr to
  determine the absolute motion.  We have not determined offsets for
  our own sample of reference stars, but given the agreement of our
  relative measurements with \citealp{lep05} we apply the same
  offsets.  The only definitive way to overcome this systematic effect
  for our purposes would be to simultaneously measure the proper
  motion of HR 8799 itself (which has accurate Hipparcos
  motions), however it is badly saturated in the POSS images rendering
  measurements of its position uncertain to several pixels.  We
  attempted to measure the proper motion of HR 8799 by determining
  best fit intersection of the horizontal and vertical diffraction
  spikes, obtaining a very uncertain estimate ($\mu_{RA}, \mu_{DEC}$)
  = (110, -43).  This is close to the Hipparcos value of HR 8799's
  motion: ($\mu_{RA}, \mu_{DEC}$) = ($107.88\pm0.75$, $-50.08\pm0.64$)
  mas/yr \cite{hipp97}.

Though we did identify several faint candidate objects on the POSS images with no obvious
 counterparts in NOMAD, only one appears worthy of further
 investigation with respect to the goals of this paper.  This object
 is at a separation of 4.6$\arcmin$ from HR 8799A, with proper motions
 of $(\mu_{RA}, \mu_{DEC}) = (105 \pm 34, -6 \pm 34)$.  It is a very
 faint object, detected in the POSS images with $SNR \approx 10$. In the Sloan Digital Sky Survey (SDSS) database it is
 designated J230716.69+210509.1.  Through the SDSS we also located it
 in the USNO-B1 catalog \cite{usno03}.  In USNO-B1 it is not given a
 proper motion measurement.  It quite clearly moves between POSS images
 relative to nearby background stars, and it is close to NLTT 55870
 (discussed above), which gives confidence that our detection of
 motion is not due to a local data reduction artifact.  It's
 $\mu_{DEC}$ is $1.3\sigma$ from the expected value of a common proper
 motion companion of HR 8799A, however its proximity warrants further
 attention.  We discuss this object in more detail below in section 4.1.

\section{Discussion}

 The large mass of the three planets combined implies a very large
initial stellar nebula, and their large orbital radii present
challenges for both the core accretion \citep{for08} and the
gravitational instability theories of planet formation
\citep{bos97}. This system could perhaps be our best evidence for a
new ``non-core-accretion'' mode of planet formation -- hence the
detection of any additional companions (planetary or higher in mass)
is very important.

The observations of Marois et al. raise several interesting questions.
The photometry shows a conspicuous lack of absorption by Methane
redwards of 1.6 \(\mu\)m, which is expected for such relatively cool
($T_{eff}\sim870$ K) objects like HR 8799b --by all the ``hot-start''
\citep{bar02, bur03} or ``core-accretion'' \citep{for08} synthetic
spectra models.  Having performed photometry at several bands, they
attempted to fit spectral energy distributions (SEDs) generated by a
hybrid atmospheric modeling code.  The best fit SEDs produced
effective temperatures over 1400K (\cite{mar08} on-line
supplement).  The
authors argue that such high temperatures cannot be supported by the
observed low luminosities unless the objects are unreasonably small or
each has significant dust extinction and reddening. They claim 3
edge-on dust disks --which would be misaligned with the plane of their
orbits (which is nearly face-on)-- could explain such reddening, and
so they rejected this dust reddening idea.

We find that HR 8799 appears to be reddened by no more than
$A_H\sim0.3$ mag (from 2MASS colors in \cite{mar08} compared to a ZAMS
A5), making it impossible that line of sight extinction to HR 8799A
alone is the cause. However, the lack of a ``reverse'' parallax in the
astrometric residuals of b's separation from A (see Fig. \ref{fig5})
offers strong proof that HR8799b is a physical companion of A. Hence,
it is unclear why HR8799b appears underluminous for its best fit model
temperatures. Perhaps strongly non-LTE effects suppress $CH_4$ in the
outer atmosphere \cite{mar08}. In any case, the discovery of other
wide reddened (or underluminous) companions would be very interesting in
this system. Therefore we need to investigate the properties of each
proper motion candidate.

\subsection{Individual Objects with Similar Proper Motion to HR 8799}

The NOMAD database contains 15 objects with similar ($\pm 20$ mas/yr)
proper motion to HR 8799A. NOMAD \#1115-0634383 (see Table
\ref{tbl-2}) has a wide $48\arcmin$ separation from HR 8799A. It is
red, with V = 16.64 and B-V color of 1.61 from the NOMAD magnitudes.
NOMAD \#1108-0634609 is a Tycho 2 star, TYC 1717-1120-1.  It is
noteworthy due to its relatively bright V = 10.62.  Though very likely
not gravitationally bound to HR 8799A (since at projected separations
$>1$x$10^5$ AU from HR 8799A they would be wider than the widest
known binaries (\citealp{clo90}; \citealp{clo08})), these objects may
warrant further study to determine if they are co-moving with the HR
8799 system.  The other objects in Table \ref{tbl-2} appear to be a
combination of too faint and too widely separated to warrant such
speculation.

Our 1 square degree manual search of the POSS images around HR8799
discovered one nearby object within $1.3\sigma$ of HR8799.  We will briefly discus this ``common
proper motion'' candidate (SDSS J230716.69+210509.1) noted in Table
\ref{tbl-3}.  We converted the SDSS photometry using the relations
given by \citealp{jga06}, obtaining $V=20.83\pm0.12$ mag and
$B-V=1.78\pm0.10$ mag.  It is not in the 2MASS point source catalog,
however there are $\approx4\sigma$ detections just above the
non-Gaussian noise in the J and Ks 2MASS images.

Though our measurement of its proper motion is marginally similar to
HR 8799, it appears to be too faint/blue to have a physical association --based on its colors and magnitude it may be more distant
than 39.9 pc.  We by no means claim that this is definitive, as color
based parallaxes are very uncertain. Due to its relative proximity to
HR8799 and the high interest in this system, further effort to confirm
our proper motion measurement, obtain IR photometry, and perhaps
obtain a radial velocity is warranted.

In summary, a cursory inspection of the proper motions and available
photometry for these objects yields no strong candidates for a bound stellar
companion to HR 8799.  Nonetheless, given the importance of the HR 8799 system
and the many astrophysical puzzles it presents, at least these 3 objects
warrant further scrutiny.

  There is still a need for future work in the search for wide
companions to HR 8799. Due to the brightness of HR 8799 on survey
images there exists an annulus from $\approx15\arcsec < r <
\approx60\arcsec$ which has not been adequately explored in this paper
for stellar companions --let alone planetary mass objects.  The
discovery of any such objects in this annulus could provide new
insights into the planetary companions of HR 8799.



\section{Conclusions}

We have made the first direct comparison of newly WiFi ADI reduced
archival Gemini AO images to archival HST/NICMOS coronagraphic
images. With 9 years between these epochs we can accurately assess the
proper motion nature of each companion. We find that both objects are
unbound to HR 8799 and are background. In this paper our major
conclusions are:

1. We estimate that no bound companions of H$<22$ mag exist from
$\sim5-15$\arcsec  from HR 8799A.

2. Any unseen scattered giant extrasolar planets in the HR 8799 system
are $>600$ AU and/or less than $\sim3 M_{Jup}$ in mass.

3. The residuals in the current published astrometry of HR 8799b's
   orbit are not explained by forcing HR 8799b to be a background
   object at $>100$ pc.

4. While we identify no clearly bound companions to HR 8799A in our
images (beyond the extrasolar planet HR 8799b), our search yields 16 objects
within 1 degree in the USNO catalog or POSS plates with similar
($\pm20$ mas/yr) proper motions to HR 8799A. Three of which merit
follow-up observations.

\acknowledgments

The authors would like to thank Glenn Schneider for helpful
discussions about NICMOS coronagraphic data and Andy Skemer for
helpful discussions about the LSR. This paper utilized data from the
HST and Gemini archives. This paper extensively used IRAF, the DSS,
2MASS, SDSS and Simbad databases. LMC is supported by an NSF CAREER
award and JRM is supported by the 2008 Steward Observatory Graduate
Fellowship.



{\it Facilities:} \facility{Gemmini}, \facility{HST (NICMOS)}.

\clearpage



\clearpage

\clearpage

\clearpage

\clearpage

\clearpage

\clearpage

\clearpage

\clearpage

\clearpage

\clearpage

\begin{table}
\begin{center}
\caption{Relative Astrometry of the 14\arcsec Companions w.r.t. HR
8799A\label{tbl-1}}
\begin{tabular}{crrrrrrrr}
\tableline\tableline
Comp. & Epoch & H  & $\Delta RA$ & $\Delta DEC$ & Sep. & P.A. & ref.\\
name & (UT) & mag & (arcsec) & (arcsec) & (arcsec) & (deg) & &\\ 
\tableline
``B'' &Oct. 30, 1998& 21.6 & $3.756\pm0.09$  & $13.19\pm0.09$ & $13.71\pm0.08$\tablenotemark{a} & $15.9\pm0.1$ & low05\tablenotemark{b}\\
``B'' &Oct. 25, 2007& NA & $3.24\pm0.02$ & $13.78\pm0.07$ & $14.15\pm0.07$ & $13.24\pm0.07$ & new\tablenotemark{c}\\
\tableline
``C'' &Oct. 30, 1998& 20.4 & $14.29\pm0.09$ & $-6.45\pm0.09$ & $15.68\pm0.08$\tablenotemark{a} & $114.3\pm0.1$ & low05\tablenotemark{b}\\
``C'' &Oct. 25, 2007& NA & $13.42\pm0.07$ & $-6.40\pm0.03$ & $14.86\pm0.07$ & $115.5\pm0.1$ & new\tablenotemark{c}\\    

\tableline
\end{tabular}

\tablenotetext{a}{The HR 8799A to B (or C) separation is as given by
\cite{low05}, the errors are as given by \cite{low05}.}
\tablenotetext{b}{data from \cite{low05}.} 
\tablenotetext{c}{data from
our WiFi ADI reduction, astrometric errors dominated by 0.5\%
platescale errors across the field. }

\end{center}
\end{table}

\clearpage

\begin{table}
\begin{center}
\caption{NOMAD Objects within $\pm1^o$ With Similar ($\pm20$ mas/yr)
Proper Motions to HR 8799\label{tbl-2}}
\begin{tabular}{crrrrrrr}
\tableline\tableline
NOMAD ID & SEP &$\mu_{RA}$&$\mu_{DEC}$&B&V&R\\
 & (\arcmin)&mas/yr&mas/yr&mag&mag&mag\\
\tableline
1113-0634282 & 33.4 & 112.0 & -64.0 &  20.41 &   ... &   19.61\\
1104-0645427 & 42.8 & 106.0 & -52.5 &  18.35 & 17.97 &   ...\\
1115-0633369 & 43.0 & 120.0 & -46.0 &  21.53 &  ...  &  19.50\\
1111-0630350 & 43.4 & 124.0 & -58.0 &  19.87 &  ...  &  19.76\\
1104-0645407 & 43.9 &  90.0 & -56.0 &  22.00 &  ...  &   ...\\
1107-0632334 & 44.6 & 104.0 & -34.0 &  18.25 & 17.62 &  16.87\\
1104-0645709 & 45.2 & 116.0 & -54.0 &  21.46 &  ...  &  20.25\\
1115-0634383\tablenotemark{a} & 48.4 & 108.0 & -60.0 &  18.25 & 16.64 &  15.91\\
1118-0648887 & 48.7 &  96.0 & -50.0 &  21.01 &  ...  &  19.93\\
1119-0648440 & 53.7 &  94.0 & -50.0 &  17.94 &  ...  &  19.71\\
1117-0654704 & 55.2 & 120.0 & -54.0 &  21.81 &  ...  &  20.13\\
1114-0630617 & 56.5 &  96.0 & -52.0 &  20.95 &  ...  &  20.09\\
1111-0631888 & 57.9 &  96.0 & -50.0 &  21.12 &  ...  &  19.37\\
1103-0652623 & 58.9 & 108.0 & -50.0 &  20.92 &  ...  &  19.86\\
1108-0634609\tablenotemark{a}\tablenotemark{b} & 59.2 & 124.4 & -51.4 &  11.61 & 10.62 &  10.02\\
\tableline
\end{tabular}
\tablecomments{Table \ref{tbl-2}: V and R magnitudes are not always given in NOMAD. The proper motion of HR 8799 is 107.88 and -50.08 mas/yr.}
\tablenotetext{a}{We discuss this object in section 4.1.}
\tablenotetext{b}{Also known as TYC 1717-1120-1.}
\end{center}
\end{table}

\clearpage

\begin{table}
\begin{center}
\caption{Faint Object 280\arcsec from A with Similar Proper Motions From
POSS\label{tbl-3}}
\begin{tabular}{crrrrrr}
\tableline\tableline
SDSS \#  & SEP  & $\mu_{RA}$ & $\mu_{DEC}$ & V & B-V\\
    & (\arcmin)&   mas/yr      &      mas/yr    &mag&mag\\
\tableline 
J230716.69+210509.1\tablenotemark{a}&4.6&105$\pm34$ mas/yr&-6$\pm34$ mas/yr&$20.83\pm0.12$&$1.78\pm0.10$ \\
\tableline
\end{tabular}
\tablecomments{Table \ref{tbl-3}: Proper motion vector as measured from POSS plates in this work.
V magnitude, B-V color, and associated uncertainties were derived
from the SDSS photometry using the formulas of \citealp{jga06}. The proper motion of HR 8799 is 107.88 and -50.08 mas/yr.}
\tablenotetext{a}{Also known as USNO-B1 1110-0590705.}
\end{center}
\end{table}




\end{document}